# Global self-similar scaling of terrestrial carbon with aridity

Jun Yin[1,2], Amilcare Porporato[3,4]

[1]Key Laboratory of Hydrometeorological Disaster Mechanism and Warning of Ministry of Water Resources; Nanjing University of Information Science and Technology, Nanjing, 210044, China
[2]Department of Hydrometeorology, Nanjing University of Information Science and Technology; Nanjing, 210044, China
[3]Department of Civil and Environmental Engineering, Princeton University; Princeton, New Jersey, 08544, USA.
[4]High Meadows Environmental Institute, Princeton University; Princeton, New Jersey, 08544, USA.)

While it is well known that water availability controls vegetation growth and soil microbial activity, how aridity affects ecosystem carbon patterns is not completely understood. Towards a more quantitative assessment of terrestrial carbon stocks, here we uncover a remarkable self-similar behavior of the global carbon stock. Using international survey and remote sensing data, we find that the key statistics (e.g., mean, quantiles, and standard deviation) of carbon stock tend to scale with the hydrological regime (i.e., aridity) via a universal exponent. As a result, when normalized by its averages in the corresponding hydrological regime, the carbon stock distributions collapse onto a single curve. Such a scaling reflects the strong coupling between hydrological cycle and biogeochemical process and enables robust predictions of carbon stocks as a function of aridity only.

# Introduction

Due to its critical role in supporting cellular structure, water is a necessary ingredient for life (*1*). As a result, the amount of water in the soil controls the terrestrial stock of soil organic carbon and biomass (*2*, *3*). Dry climates with lower plant productivity are prone to land degradation and are threatened in terms of biodiversity, while wet environment with fertile soils have potential shift from carbon sinks to sources in changing climates (*4–6*). Water availability is expected to change in response to global warming (*7*) but the projections remain uncertain (*8*, *9*). Such uncertainties propagate into the global biogeochemical cycles, posing great challenges not only for high-resolution Earth system modeling (*10*, *11*) but also for nature-based approaches to climate-change mitigation (*12*, *13*).

Aridity controls on ecosystem processes have received significant research attention recently, highlighting the nonlinear dependence of the biogeochemical process and plant dynamics on water availability (*14–17*). However, such studies typically focused on a limited range of aridity for specific types of ecosystems (see a brief review of these studies in Table S1) or only analyzed average ecosystem attributes without exploring their full statistical distributions. Extending the analysis to the entire spectrum of aridity, from wetlands to forests and deserts, is crucial to achieve robust estimations of global carbon stocks and predict their future distributions.

Toward this goal, here we used global survey and remote sensing data sets to systematically quantify the impacts of water availability on carbon stocks of terrestrial ecosystems. We found the self-similar scaling of carbon stocks with aridity and its universal distribution across all ecosystems, highlighting the critical control of hydrological process on global carbon cycle and its potential impacts on the design of climate mitigation policy.

# Results

We began by analyzing the dryness index, a well-known aridity indicator, defined as the ratio of long-term averages of potential evaporation, $E_{max}$, to precipitation, $R$,

$$D_I = \frac{E_{max}}{R}. \tag{1}$$

This dimensionless index is historically used in hydrology for partitioning precipitation into evaporation and runoff (*18–20*) and is also widely used in ecology to characterize ecosystem functioning (*17*, *21*, *22*). As shown in Figure 1, even by simple inspection, the global distribution of dryness index already shows great similarity with that of total carbon stocks defined as the sum of biomass and soil organic carbon (see Methods). The wet and carbon-rich regions are located in the tropics and at high latitudes, such as the Amazon rainforest, Indonesia, and the peatlands in Canada and Siberia; as well known, deserts such as Sahara and Atacama, some of the driest regions in the world, are also lowest in carbon content. While the carbon stock is also influenced by various factors associated with photosynthesis and soil respiration processes, these biotic and abiotic factors are highly interconnected and associated with the local aridity (see Fig. S1), thus leading to a dominant control of dryness index on carbon stocks.

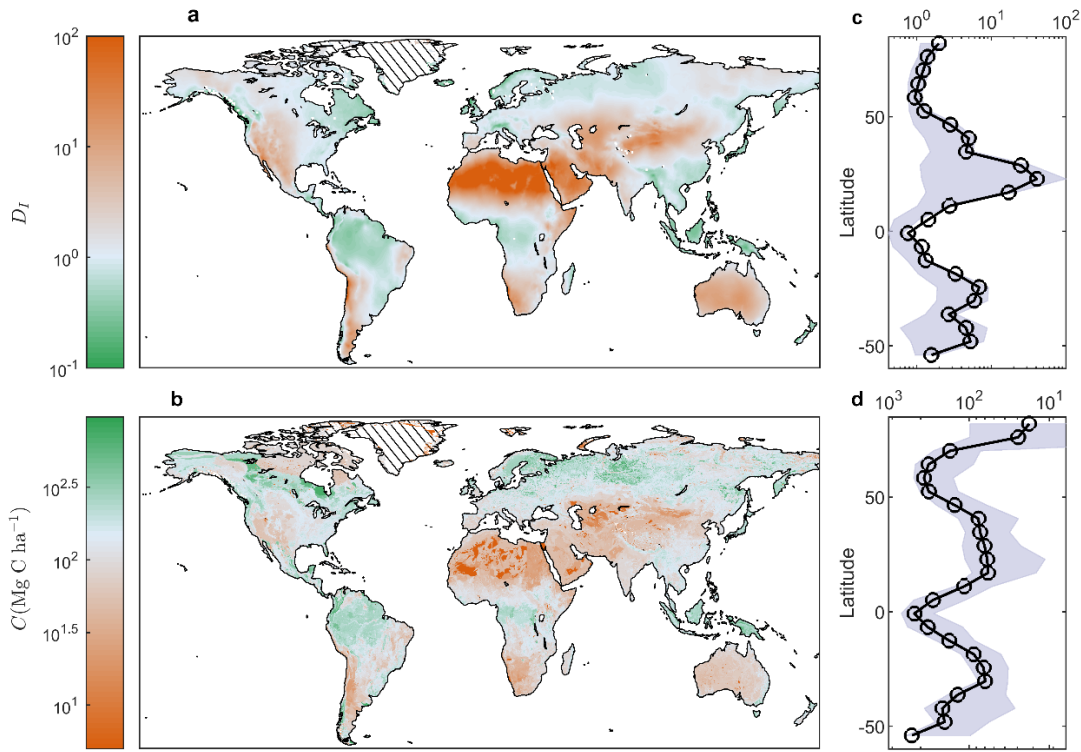

Figure 1 Comparison between dryness index and total carbon stocks. (**a**) Global distributions of dryness index, $D_I$, calculated from Eq. (1) with data from Climate Research Unit (CRU) TS version 4.04. (**b**) Global distributions of total carbon stocks, where the data is from Harmonized global maps of above and belowground biomass *(23)* and Harmonized World Soil Database *(24)*. The right-side plots show the latitudinal averages (circles) and 1st-3rd quartiles (shaded areas) of (**c**) dryness index and (**d**) total carbon stocks. Lands with water/ice and urban lands (hatched areas) are excluded from this study; color scheme and data are presented in logarithmic scales.

To precisely quantify this carbon-aridity relationship, we analyzed the carbon distributions as a function of the dryness index (see Methods). Remarkably, the results show that, within each hydrological regime (i.e., conditional on dryness index), the average total carbon, $\mu_C$, can be well approximated as a power-law function of dryness index, $D_I$ (see Figure 2)

$$\mu_C = \mu_0 D_I^b \qquad (2)$$

where the exponent $b = -0.4$ and $\mu_0 = 145$ Mg ha$^{-1}$ (confidence intervals and the corresponding statistics are given in Methods). This global scaling may be understood as the combined effect of vegetation and soil microbial response to water availability. On the one hand, the increase in carbon input through photosynthesis increases with soil moisture, but tends to plateau in wet conditions *(25)*. On the other hand, the carbon loss due to decompositions is linked to soil microbial activity, which instead declines with anoxic conditions, thus favoring the accumulation of carbon *(26)*.

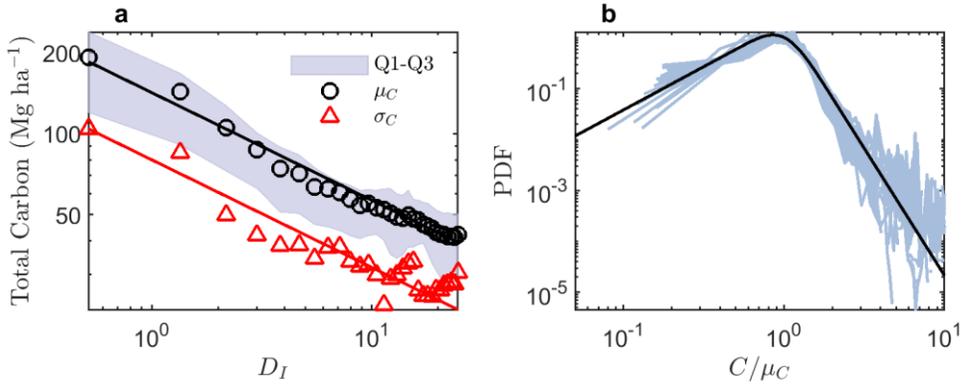

Figure 2 Self-similar behavior of global carbon stocks. (**a**) Relationship between carbon stocks and aridity. For equal dryness-index intervals, circles are the corresponding mean value of carbon stocks (sum of biomass and soil organic matter), triangles are the standard deviations, the shading areas show the upper and lower quartiles, and black and red lines are the power-law fits of the means and standard deviations. (**b**) Normalized distributions of carbon stock from different hydrological regimes (Fig. S2 presents these distributions separately). The thick black curve is the fitted Double Pareto Lognormal distribution and the corresponding quartile-quartile plots are given in Fig. S3. Data sources are provided in Methods.

A second major finding of the statistical analysis was that the variability of the carbon stock distributions (see the shaded area in Figure 2a) also shows a scaling behavior similar to the mean. Specifically, the standard deviation in each hydrological regime ($\sigma_C$, see red triangles in Figure 2a) is well represented by

$$\sigma_C = \sigma_0 D_I^b \qquad (3)$$

where $\sigma_0 = 76$ Mg ha$^{-1}$ and the exponent is basically the same as for the mean, $b = -0.4$ (confidence intervals are given in Methods). Thus, while the standard deviation decreases with the dryness index, the coefficient of variation, obtained by dividing Eq. (3) by (2), is approximately a constant, equal to 0.55, across `all terrestrial ecosystems. This signifies that the relative dispersion of carbon stocks is invariant in each hydrological regime.

This global pattern also seems to be invariant across different spatial scales. While, as expected, upscaling grid data reduces the spatial heterogeneity and thus lowers the variances of carbon in each hydrological regime (see red triangles in Fig. S4 a-c), the corresponding mean carbon stocks still scale with dryness index with nearly identical exponent for data at different spatial resolutions (see black circles in Fig. S4 a-c). Even in the extreme case of lumping the data across different Köppen-Geiger climate zones, this power-law relationship remains valid (see Fig. S4d), pointing to the robustness of this self-similar scaling behavior.

Looking at the full distributions of carbon stocks in each hydrological regime a further remarkable result emerges: when normalized by their averages ($c = C/\mu_C$), these distributions collapse onto a single curve with two distinct power-law tails (see Figure 2b). Such a curve is well fitted well by a Double Pareto Lognormal distribution (see Methods). The collapse of the distributional tails, with their algebraic (i.e., not exponential) decay, thus reveals another self-similar behavior, which is indicative of a wide distribution with large probabilities of both extremely high or low carbon stocks, regardless of the ecosystem type and hydrologic regime. Such a universal behavior is noteworthy, especially in view of the many factors controlling the soil-plant carbon dynamics, and points to a common, fundamental way in which regional heterogeneities in aridity impact the local carbon stocks.

Besides the global soil survey data used in above analysis, we also checked the scaling behavior in other model-based soil mapping data previously used in the literature. These datasets use state-of-the-art machine learning models to link soil carbon as functions of climate, landscape, land use/cover, and other related variables, and also include soil carbon in deeper soils (see Methods). The results show that the power-law scaling is still valid, but with slight deviations in the extreme wet or dry regimes (see Fig. S5-7). The distribution of the normalized carbon stocks again collapses onto a single curve although the upper tails seem to have a somewhat faster decay (i.e. exponential). It should be noted that the carbon-aridity relationship obtained from these model-based datasets reflects the features of the training samples (which do not cover the entire globe) and could be influenced by model structures.

## Discussion

The global scaling may have significant practical implications. Given the strong controls of aridity on soil nutrient availability, plant growth, and carbon fluxes, identified by $CO_2$ fertilization experiments, long-term observations, and large-scale modeling (*27–30*), as well as its robustness across climates, it is logical to surmise that the global carbon-aridity relationships discovered here will also hold in future climates. If this is the case, expressing the carbon stock distribution conditional on hydrological regime as a function of the dryness index would allow us to estimate trends of future global carbon stocks only based on the dryness index projections. Such an approach would complement global model predictions , which themselves are affected by uncertainties linked to extrapolations of model results and often tend to be linked more to the type of application rather than to the spatial and temporal scales of interest (*31*).

For example, one could use the present results to speculate about possible trends in carbon storage, following typical climate scenarios. Assuming that the dryness index changes relatively slowly so that the carbon cycle evolving quasi-statically (see Methods), in an Earth-greening scenario (*32*) of 1% decrease of dryness index for all land cover, the carbon gain would be 0.4% of total terrestrial carbon stocks (i.e., $C_t (1-0.99^b)$ = 7.3 Pg C), while assuming a global drying (*33*) of the same magnitude one would have approximately the same amount of carbon loss.

In conclusion, the global scaling of carbon stocks unveiled here allows us to link aridity projections to soil-plant carbon stocks, providing quantitative estimates of carbon storage trends for ecosystem attaining different aridity conditions. While of course the details of future carbon storage are also related to the speed at which such different dryness indices are achieved, these quasi-static estimates may be useful to inform climate-mitigation strategies. Future work should focus on the genesis and justification of the global self-similarity behavior found as well as on the effects of dynamic changes in dryness index on transient changes in carbon storage; upscaling these dynamics, from the pulsing dynamics at the local scales to the global and multi-decadal scales of climatic interest (*34*), remains a crucial open problem for future carbon-storage predictions.

**Materials and Methods**

Data and Data Availability

We used the monthly potential evapotranspiration and precipitation data from Climate Research Unit (CRU TS v 4.04, https://crudata.uea.ac.uk/cru/data/hrg/), which is a widely used climate dataset obtained by interpolating climate variables from global networks of weather station observations (*35*). The dryness index was calculated as the ratio of the long-term averages of potential evapotranspiration and precipitation as defined in Eq. (1).

To find the global carbon stocks, we summed the harmonized global maps of above and belowground biomass (HGMB, doi.org/10.3334/ORNLDAAC/1763) and Harmonized World Soil Database (HWSD, https://daac.ornl.gov/SOILS/guides/HWSD.html). The former is the most up-to-date biomass datasets integrated from several remote sensing datasets and other ancillary maps (*23*); the latter is one of the most complete soil datasets developed by Food and Agriculture Organization in collaboration with multiple international soil research centers (*24*). The global biomass is estimated to be 409 Pg C and global soil carbon is 1417 Pg C, for a total of $C_t$ = 1826 Pg C global terrestrial carbon stocks. Note that HWSD estimates soil carbon stocks only up to 1 m depth of soil, which may neglect certain deep soil carbon, for example, in some peatlands (*36*, *37*). We also use a model-based dataset to account for carbon stocks up to 2 m soil (*38*) and the IPCC Tier-1 Global Biomass Carbon (*39*) to test the carbon-aridity relationship from different data sources; these data are available at (github.com/whrc/Soil-Carbon-Debt; cdiac.ess-dive.lbl.gov/epubs/ndp/global_carbon/carbon_documentation.html)

The resolutions for CRU, HGMB, HWSD are 0.5 degree, 10 arc-second, and 30 arc-second, respectively. These data were interpolated into the same equal-area grids with 0.05 degree resolution in the equator and uniform latitudinal spacing (*40*). This interpolation to equal-area grids not only provides pairwise datasets for quantifying carbon-aridity relationship, but also facilitates the estimation of distributions and quantiles of carbon stocks without treating the different geodetic zonal weights in any grid points.

In our analysis of the carbon-aridity relationship, we also excluded urban and build-up land which may subject to significant anthropogenic perturbation; we did not account for lands with snow/ice/water cover due to unavailable soil information (see hatched area in Figure 1). This results in a total of 3720647 grid points. The land cover types were identified from Moderate Resolution Imaging Spectroradiometer (MODIS) Land Cover Climate Modeling Grid Product (modis.gsfc.nasa.gov/data/dataprod/mod12.php).

Carbon Statistics Conditional on Dryness Index

We used the pairwise carbon-aridity datasets with equal-area grids to quantify the conditional statistics of carbon stocks. We classified 30 hydrological regimes by dryness index ranging from 0.1 to 25 with equal interval. We found the corresponding statistics (first and third quantiles, averages, and standard deviation) in each regime, which were then compared with the dryness index (see Figure 2a). Different classifications with variable intervals over an extended range of aridity did not significantly change our results (e.g., see Fig. S8).

We used least square method to fit the conditional averages and standard deviation of carbon stocks with dryness index using the power-law expressions in Eqs. (2) and (3). The calibrated parameters are $\mu_0$ = 146 (141.8, 150.6) Mg ha$^{-1}$, $b$ = -0.4 (-0.44, -0.40) with coefficient of determination of 0.98, and $\sigma_0 = 78.9$ (73.4, 84.4) Mg ha$^{-1}$, $b = -0.4$ (-0.43, -0.35) with coefficient of determination of 0.91, where the brackets indicate the 95% confidence bounds. The parameters $b$ in these two expressions are identical, when rounded to one decimal figure.

Distributions of Carbon Stocks

The Double Pareto Lognormal distribution has been used in economics and physical sciences (*41*) and its double power-law tails nicely fits the normalized distribution of global carbon stocks, $c = C / \mu_C$. Its probability density can be expressed as (see Figure 2b)

$$f_n(c) = \frac{\alpha\beta}{\alpha+\beta} c^{-\alpha-1} \exp\left[\alpha v + \alpha^2 \tau^2 / 2\right] P_n\left(\frac{\ln c - v - \alpha\tau^2}{\tau}\right) + \frac{\alpha\beta}{\alpha+\beta} c^{\beta-1} \exp\left[-\beta v + \alpha^2 \tau^2 / 2\right] \left[1 - P_n\left(\frac{\ln c - v + \beta\tau^2}{\tau}\right)\right], \quad (4)$$

where $P_n()$ is cumulative distribution function of standard normal distribution, $\alpha$ and $\beta$ define the upper and lower tails, $v$ and $\tau$ define the lognormal distribution in the body part. This distribution has fitted parameters $\alpha = 4$, $\beta = 2.7$, $v = -0.01$, and $\tau = 0.2$ and was graphically compared in the quantile-quantile plots in Fig. S3. Combining this normalized distribution, with Eq. (2), we can obtain the distribution of carbon stock conditional on dryness index as a derived distribution, i.e., $\mu_0^{-1} D_I^{-b} f_n(\mu_0^{-1} D_I^{-b} C)$.

Changes in Carbon Stocks

To evaluate climate change impacts on carbon stocks, we assume that the power-law carbon-aridity relationships in Eq. (2) is robust to future conditions under the quasi-steady state assumption. Accordingly, when a region changes dryness index by a factor $k$, the expected carbon stocks will be scaled with $k^b$. At global scale the expected change of carbon stocks can be expressed as

$$\Delta C = C_t (1 - k^b), \quad (5)$$

where the total terrestrial carbon stocks $C_t$ is estimated to be 1826 Pg C from HWSD and HGMB datasets.

**Supplementary Materials**

Figs. S1 to S4

Tables S1

# Supplementary Materials for

**Global self-similar scaling of terrestrial carbon with aridity**

Jun Yin, Amilcare Porporato.

Table S1 Brief review of recent studies *(15, 17, 21, 22, 42–46)* on the explicit relationships between aridity and ecosystem attributes. Aridity ranges from 0.1 to 25 in Figure 2 and from 0.1 to 100 in Fig. S8. Note that there are various definitions of aridity used in literature, such as wetness index and aridity index, which are all converted to dryness index as defined in Eq. (1) for comparison.

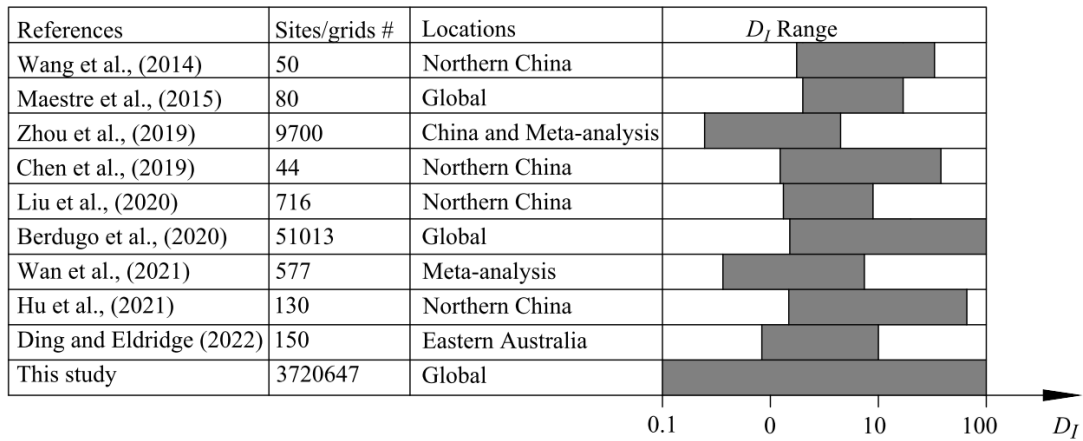

| References | Sites/grids # | Locations | $D_I$ Range |
|---|---|---|---|
| Wang et al., (2014) | 50 | Northern China | |
| Maestre et al., (2015) | 80 | Global | |
| Zhou et al., (2019) | 9700 | China and Meta-analysis | |
| Chen et al., (2019) | 44 | Northern China | |
| Liu et al., (2020) | 716 | Northern China | |
| Berdugo et al., (2020) | 51013 | Global | |
| Wan et al., (2021) | 577 | Meta-analysis | |
| Hu et al., (2021) | 130 | Northern China | |
| Ding and Eldridge (2022) | 150 | Eastern Australia | |
| This study | 3720647 | Global | |

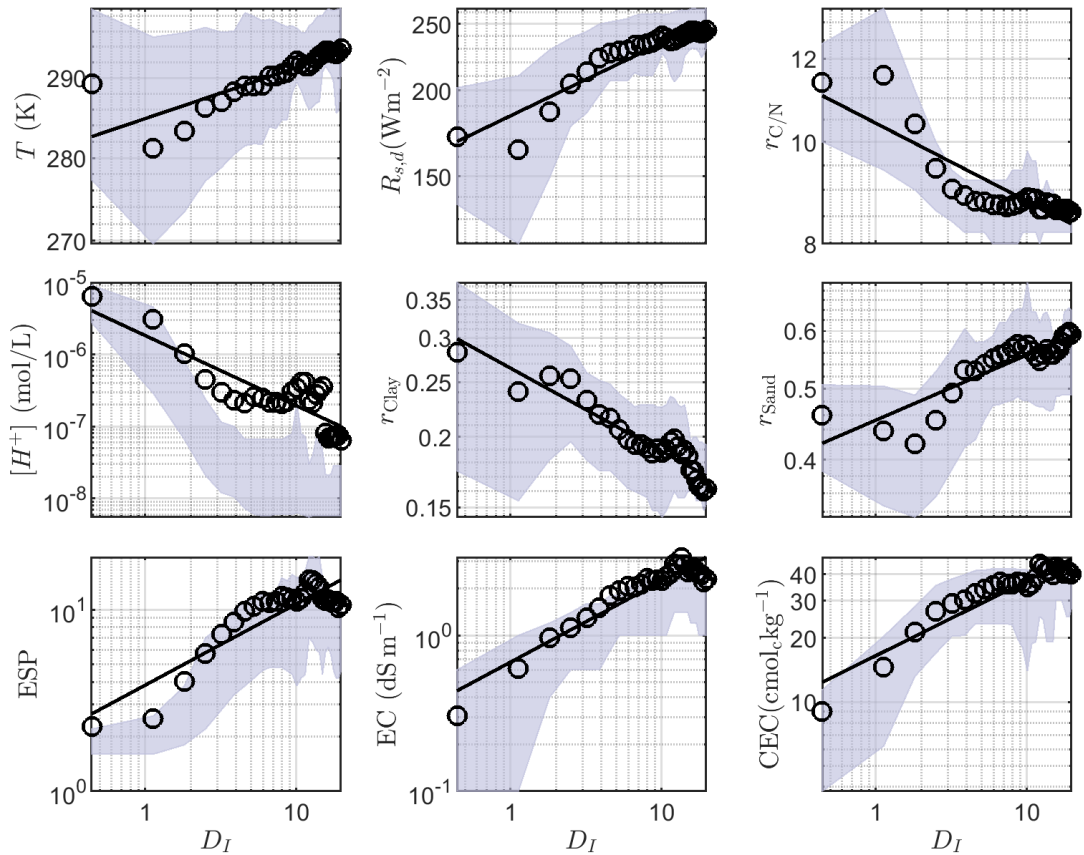

Fig. S1 Global control of dryness index on factors impacting soil carbon. In the each hydrological regime over the world (equal interval of dryness index, see Methods), the circles are the corresponding mean values, the shading areas show the upper and lower quartiles, and the lines are the power-law fits. Soil attribute data are from World Inventory of Soil property Estimates (https://www.isric.org/explore/wise-databases).

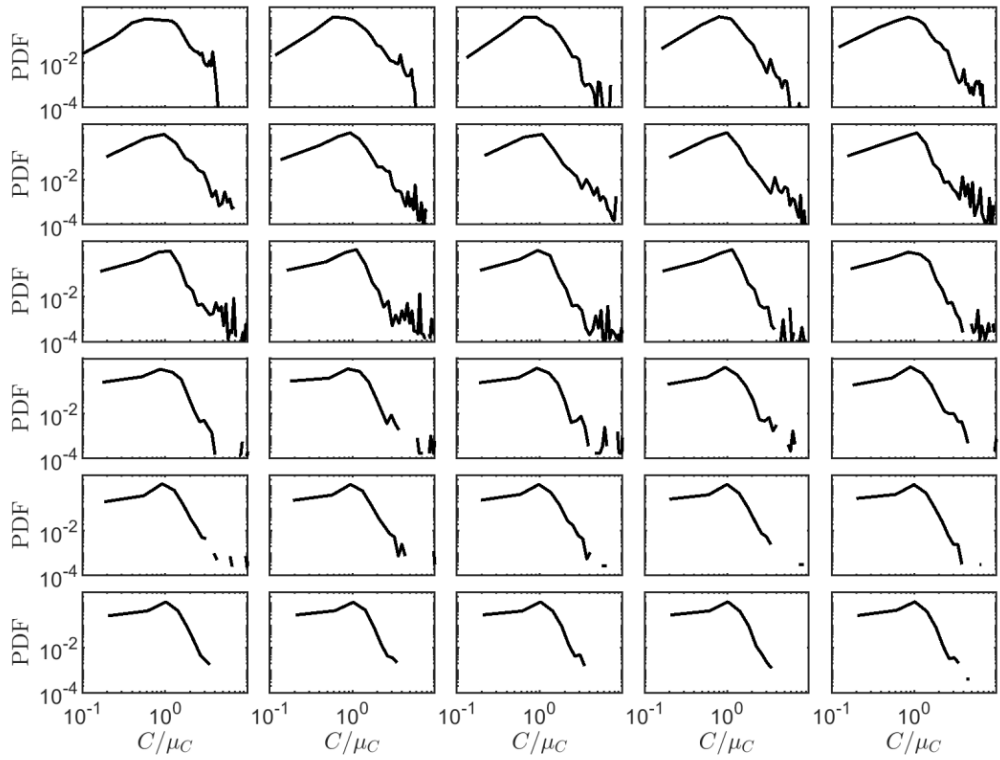

Fig. S2 As in Figure 2b in the main text but separately presented for each hydrological regime. From top left to bottom right panels, the dryness index ranging from 0.1 to 25 with equal interval of 0.83.

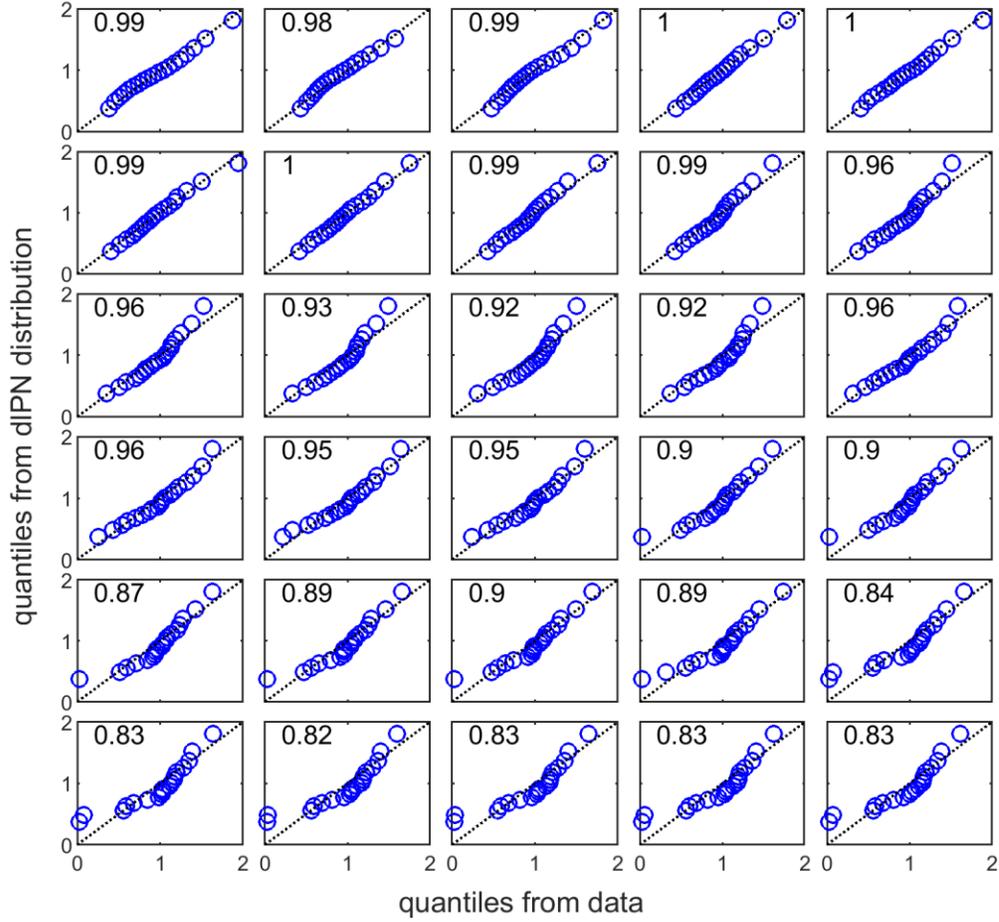

Fig. S3 Quantile-quantile plots for the Double Pareto Lognormal distributions in Figure 2b in the main text. Quantiles in the x-axis (ranging from 5 to 95 with interval 5) are from data; quantiles in the y-axis are from theoretical Double Pareto Lognormal distributions with parameters (4, 2.7, -0.01, 0.2). From top left to bottom right panels, the dryness index ranging from 0.1 to 25 with equal interval of 0.83. The value in the top right of each panel shows the coefficient of determination of the corresponding Quantile-quantile plots.

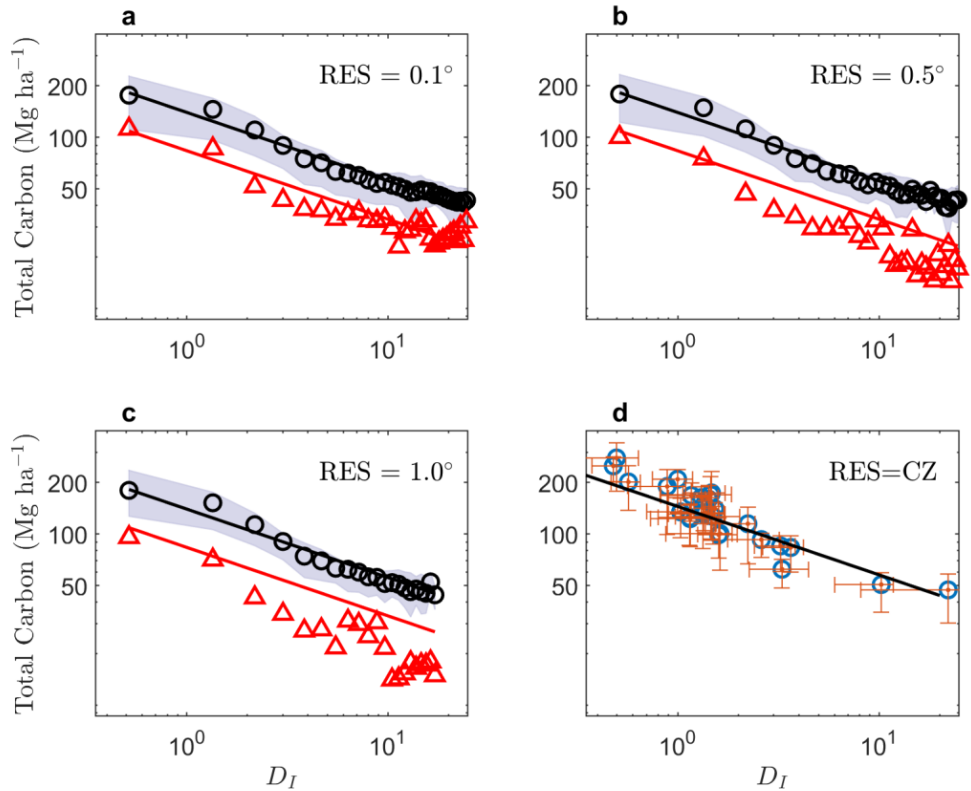

Fig. S4 As in Figure 2a in the main text but at different spatial resolutions. The carbon and aridity data were upscaled into (a) 0.2, (b) 0.5, and (c) 1.0 degree resolutions. For comparison, the black and red lines are reference lines from Figure 2a for data at 0.05 degree resolution. In panel (d), this was further upscaled to Köppen-Geiger climate zones. The horizontal and vertical error bars show 25 and 75 percentiles of the dryness index and carbon socks. Three climate zones with less than 500 grids were excluded from this analysis. See ref (*47*) for the climate zone classification.

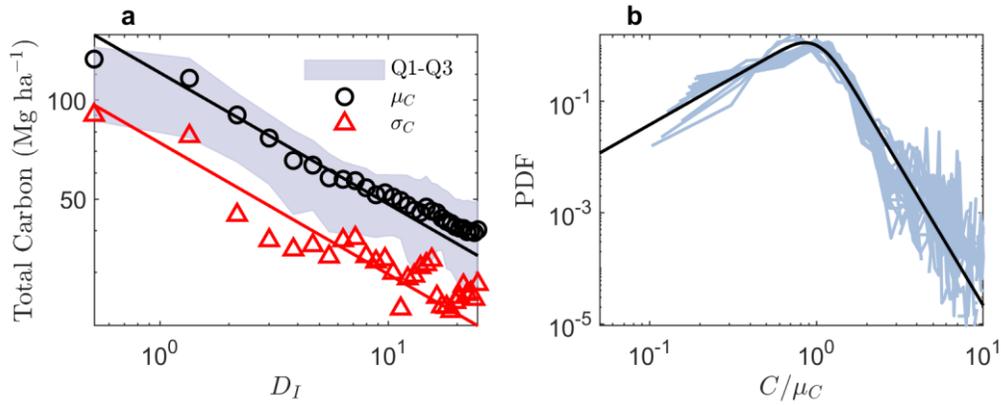

Fig. S5 As in Figure 2 in the main text but for different carbon data sources. Soil organic carbon was from HWSD and biomass was from Ruesch and Gibbs (2008) *(39)*. The calibrated parameters for (a) $\mu_0$ = 111.6 (107.7, 115.6) Mg ha$^{-1}$, $b$ = -0.33 (-0.35, -0.32) in Eq. (2); and $\sigma_0$ = 70.2 (65.0, 75.4) Mg ha$^{-1}$, $b$ = -0.36 (-0.40, -0.32) for Eq. (3). The coefficients of determination of the regression are 0.97 and 0.89, respectively. The calibrated parameters for (b) are $\alpha$ =4, $\beta$ =2.7, $v$ =-0.01, and $\tau$ =0.2.

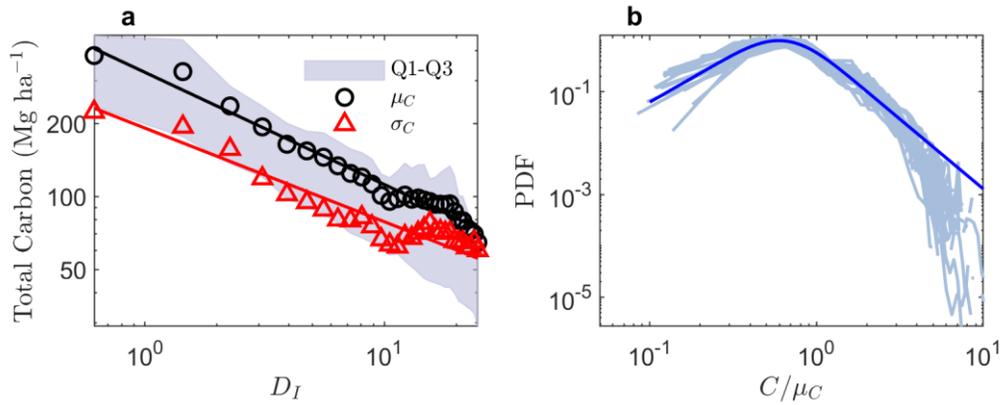

Fig. S6 As in Figure 2 in the main text but for different carbon data sources. Soil organic carbon within depth of 2m was from Sanderman et al., (2017) *(38)* and biomass was from Spawn et al. (2020) *(23)*. The calibrated parameters for (a) are $\mu_0$ = 325.1 (310.4, 339.7) Mg ha$^{-1}$, $b$ = -0.46 (-0.49, -0.43) in Eq. (2); and $\sigma_0$ = 191.4 (179.6, 203.2) Mg ha$^{-1}$, $b$ = -0.39 (-0.42, -0.35) for Eq. (3). The coefficients of determination of the regression are 0.97 and 0.93, respectively. The calibrated parameters for (b) are $\alpha$ =1.72, $\beta$ =2.9, $v$ =-0.36, and $\tau$ =0.32.

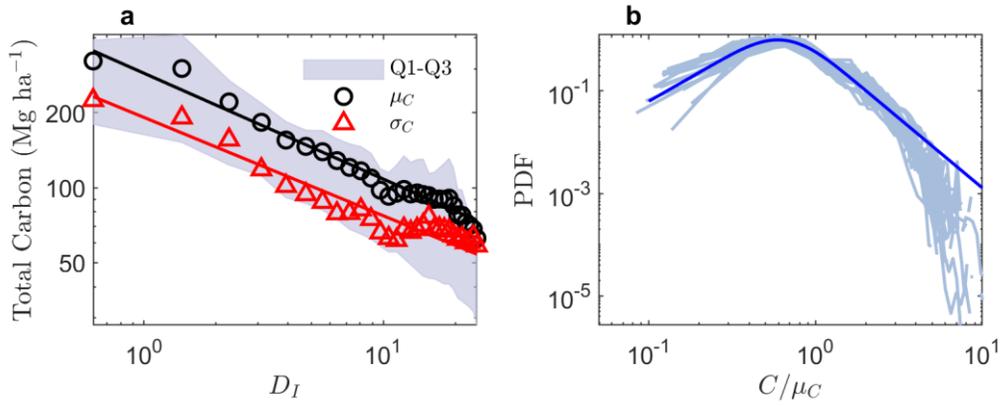

Fig. S7 As in Figure 2 in the main text but for different carbon data sources. Soil organic carbon within depth of 2m was from Sanderman et al., (2017) *(38)* and biomass was from Ruesch and Gibbs (2008) *(39)*. The calibrated parameters are $\mu_0$ = 288.2 (272.8, 303.7) Mg ha$^{-1}$, $b$ = -0.42 (-0.46, -0.39) in Eq. (2); and $\sigma_0$ = 191.6 (180.7, 202.5) Mg ha$^{-1}$, $b$ = -0.39 (-0.43, -0.36) for Eq. (3). The coefficients of determination of the regression are 0.96 and 0.94, respectively. The calibrated parameters for (b) are $\alpha$ =1.72, $\beta$ =2.9, $\nu$ =-0.36, and $\tau$ =0.32.

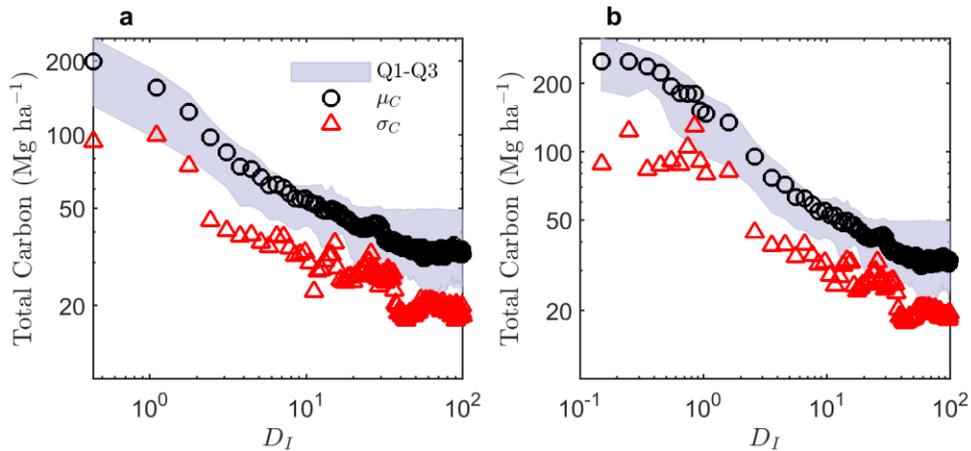

Fig. S8 As in Figure 2b in the main text but for an extend range of $D_I$ with (**a**) equal binning interval of 2/3 and (**b**) variable intervals (0.1 for $0.1 < D_I < 1$ and 1 for $D_I \geq 1$). Dense intervals within $0.1 < D_I < 1$ were used to clarify the carbon-aridity relationship in wet regions.